\documentclass[aps,prc,superscriptaddress,showpacs,nofootinbib,notitlepage]{revtex4-1} 
\usepackage{graphicx,amsmath,amssymb,bm,color}

\DeclareGraphicsExtensions{ps,eps,ps.gz,frh.eps,ml.eps}
\usepackage[english]{babel}
\usepackage{fmtcount} 
\usepackage{graphicx}
\usepackage{amsmath, amsfonts, amssymb, amsthm, amscd, amsbsy, amstext} 
\usepackage{marvosym}  
\usepackage{braket,xspace}
\usepackage{mathrsfs} 
\usepackage{booktabs}
\usepackage{upgreek}
\usepackage{array}
\usepackage[makeroom]{cancel}
\usepackage{ wasysym }
\usepackage{blindtext}
\usepackage{url}
\usepackage{hyperref}

\usepackage[shortlabels]{enumitem}

\usepackage{epigraph}
\DeclareMathOperator{\e}{\operatorname{e}}

\makeatletter
  
\newcommand{\overleftrightsmallarrow}{\mathpalette{\overarrowsmall@\leftrightarrowfill@}}
\newcommand{\overrightsmallarrow}{\mathpalette{\overarrowsmall@\rightarrowfill@}}
\newcommand{\overleftsmallarrow}{\mathpalette{\overarrowsmall@\leftarrowfill@}}
\newcommand{\overarrowsmall@}[3]{%
  \vbox{%
    \ialign{%
      ##\crcr
      #1{\smaller@style{#2}}\crcr
      \noalign{\nointerlineskip}%
      $\m@th\hfil#2#3\hfil$\crcr
    }%
  }%
}
\def\smaller@style#1{%
  \ifx#1\displaystyle\scriptstyle\else
    \ifx#1\textstyle\scriptstyle\else
      \scriptscriptstyle
    \fi
  \fi
}
\makeatother

\newcommand{\kF}{k_{\rm F}}

\def\XXint#1#2#3{{\setbox0=\hbox{$#1{#2#3}{\int}$ }
\vcenter{\hbox{$#2#3$ }}\kern-.6\wd0}}

\makeatletter

\newcommand{\Rmnum}[1]{\expandafter\@slowromancap\romannumeral #1@}
\makeatother

\makeatletter
\newcommand\opteq[1]{\mathrel{\mathpalette\opt@eq{#1}}}
\newcommand{\opt@eq}[2]{%
  \begingroup
  \sbox\z@{$#1#2$}%
  \sbox\tw@{\resizebox{!}{.5\ht\z@}{$\m@th#1($}}%
  \nonscript\hskip-\wd\tw@
  \mkern1mu
  \raisebox{-.35\ht\z@}[0pt][0pt]{\resizebox{!}{.5\ht\z@}{$\m@th#1($}}%
  \mkern-1mu
  {#2}%
  \mkern-1mu
  \raisebox{-.35\ht\z@}[0pt][0pt]{\resizebox{!}{.5\ht\z@}{$\m@th#1)$}}%
  \mkern1mu
  \nonscript\hskip-\wd\tw@
  \endgroup
}
\makeatother
\newcommand{\leoq}{\opteq{\leq}}

\makeatletter

\def\env@blscases{
  \let\@ifnextchar\new@ifnextchar
  \left.
  \def\arraystretch{1.2}
  \array{@{}l@{\quad}l@{}}
}
\makeatother

\makeatletter

\def\env@rcases{
  \let\@ifnextchar\new@ifnextchar
  \left.
  \def\arraystretch{1.2}
  \array{@{}l@{\quad}l@{}}
}
\makeatother

\hypersetup{%
    pdfsubject=Paper,
    unicode=true, 
    breaklinks=true,
     colorlinks   = true,
     linkcolor = blue,
     citecolor = blue,
     menucolor = blue,
     citecolor    = blue,
     urlcolor = blue
}

\allowdisplaybreaks
\begin{document}


\title{Constrained extrapolation problem and order-dependent mappings~\footnote{Paper prepared for the Festschrift in honor of the 60th birthday of Prof.~D.A.~Drabold.}}

\author{C. Wellenhofer}
\email{wellenhofer@theorie.ikp.physik.tu-darmstadt.de}
\affiliation{Technische Universit\"{a}t Darmstadt, Department of Physics, 64289 Darmstadt, Germany}  
\affiliation{ExtreMe Matter Institute EMMI, GSI Helmholtzzentrum f\"{u}r Schwerionenforschung GmbH, 64291 Darmstadt, Germany}  

\author{D.~R.~Phillips}
\email{phillid1@ohio.edu}
\affiliation{Department of Physics and Astronomy and Institute of Nuclear and Particle Physics, Ohio University, Athens, OH 45701, USA}

\author{A. Schwenk}
\email{schwenk@physik.tu-darmstadt.de}
\affiliation{Technische Universit\"{a}t Darmstadt, Department of Physics, 64289 Darmstadt, Germany}  
\affiliation{ExtreMe Matter Institute EMMI, GSI Helmholtzzentrum f\"{u}r Schwerionenforschung GmbH, 64291 Darmstadt, Germany}  
\affiliation{Max-Planck-Institut f\"{u}r Kernphysik, Saupfercheckweg 1, 69117 Heidelberg, Germany}

\begin{abstract}
We consider 
the problem of extrapolating the perturbation series for the dilute Fermi gas in three dimensions
to the unitary limit of infinite scattering length and into the BEC region, 
using the available strong-coupling information to constrain the extrapolation problem. 
In this constrained extrapolation problem (CEP) the goal is to find classes of 
approximants that give well converged results already for low perturbative truncation orders.
First, we show that standard Pad\'{e} and Borel methods are 
too restrictive to give satisfactory results for this CEP.
A generalization of Borel extrapolation is given by the so-called Maximum Entropy extrapolation method (MaxEnt).
However, we show that MaxEnt requires extensive elaborations to be
applicable to the dilute Fermi gas and is thus not practical for the CEP in this case.
Instead, we propose order-dependent-mapping extrapolation (ODME) as a simple, practical, and general method for the CEP.
We find that the ODME approximants for the ground-state energy of the dilute Fermi gas 
are robust with respect to changes of the mapping choice
and agree with results from quantum Monte Carlo simulations within uncertainties. 
\end{abstract}

\maketitle

\section{Introduction}
\epigraph{\textit{``It's the job that's never started as takes longest to finish.'' Sam Gamgee\\
Lord of the Rings, J.R.R.~Tolkien}\\
In recognition of Prof.~Drabold's love of Middle Earth, throughout the paper we provide inspirational quotes from a classic text.}

Consider an observable $F$ of a system characterized by a coupling $x$ and let $F$ be defined relative to the noninteracting system. 
(The ground-state energy $E/E_0$ is a canonical example.)
Quantitative analytic knowledge about $F(x)$
is in most cases restricted to its weak-coupling perturbation series up to a given order $N$, i.e.,
\begin{equation}\label{eq:pertseries}
F(x)
\stackrel{x\rightarrow 0}{\simeq}
1+\sum_{k=1}^N c_k x^k + O(x^{N+1}) \,.
\end{equation}
While it provides precise information about the behavior of $F(x)$ as ${x\rightarrow 0}$,
the perturbation series generally fails to yield viable approximations away from weak coupling.

Indeed, the perturbation series is generally a divergent asymptotic series, with factorially growing coefficients at large orders, e.g., 
${c_k \stackrel{k\rightarrow \infty}{\sim}k!}$~\cite{PhysRevLett.27.461,ZinnJustin:1980uk}.
The principle of superasymptotics~\cite{Boyd} states that for a given $x$ the accuracy of the perturbation series 
increases with increasing $N$ only for ${N<N_\text{opt}}$, where the optimal truncation order $N_\text{opt}$
is given by the largest $N$ for which ${|c_N x^N|<|c_{N-1}x^{N-1}|}$.

Resummation methods such as Pad\'e~\cite{bakerbook,benderbook} and Borel resummation~\cite{ZinnJustin:1980uk} 
can produce estimates for $F(x)$ beyond superasymptotics.  
In the case of simple idealized systems where detailed knowledge of the 
analytic structure of $F(x)$ is available, Borel methods can yield approximants
that converge to the exact $F(x)$ for 
${N\rightarrow\infty}$~\cite{doi:10.1063/1.4921155,PhysRevLett.121.130405,Costin_2019}.   
However, for realistic systems, standard resummation methods generally fail to give access to the regime of strong coupling, 
in particular if only a few of the perturbation coefficients are available.

Experiment or computational methods provide access to the behavior of $F(x)$ at strong coupling,
yielding information on the limit ${F(x\rightarrow-\infty)=\xi}$ as well as the leading
coefficients $d_k$ in the strong-coupling expansion:
\begin{equation}\label{eq:sce}
F(x)
\stackrel{x\rightarrow -\infty}{\simeq}
\xi+\sum_{k=1}^M \frac{d_k}{x^k}+ O(x^{-M-1}) \,.
\end{equation}
Weak-to-strong-coupling extrapolants can then be defined as classes of functions $F_N(x)$
that reproduce the perturbation series to order $N$ and incorporate additional strong-coupling information---where the latter in general will come with numerical uncertainties.
\emph{This defines the constrained extrapolation problem (CEP).}
Formally, the goal of the CEP is to find approximants $F_N(x)$ that converge rapidly and smoothly to the correct $F(x)$ as ${N\to\infty}$. 
In practice, $F_N(x)$ should be well converged already at low orders, since typically knowledge of the perturbation series is restricted to a low truncation order 
${N\lesssim 4-6}$ (see, e.g., Refs.~\cite{Kleinert:2001ax,Wellenhofer:2018dwh,PhysRevD.100.056019,HouDrut}).

In the present paper we consider the CEP for an unpolarized gas of spin-$1/2$ fermions interacting via short-range interactions, which are characterized by the s-wave scattering length $a_s$ at low energies.
The zero-temperature properties of the system are then determined by the single variable $x=\kF a_s$, where $\kF$ is the Fermi momentum of the system.
In this way, the dilute Fermi gas constitutes a prime example of low-energy universality, 
with relevance to a variety of systems, 
in particular ultracold atoms~\cite{RevModPhys.80.885,BCSBECtheory,RevModPhys.82.1225} and neutron matter~\cite{Schw05dEFT,Kaiser:2011cg,Gandolfi:2015jma,Tews_2017}.
The weak-coupling expansion of the ground-state energy $F=E/E_0$ was recently calculated to fourth order~\cite{Wellenhofer:2018dwh,Wellenhofer:2021eis}:\footnote{We note that in our previous paper~\cite{Wellenhofer:2020ylh} there is 
a typo in the seventh digit of $c_3$, i.e., Eq. (3) of Ref.~\cite{Wellenhofer:2020ylh} should say $c_3=0.0303089(0)$ not
$c_3=0.0303088(0)$.}
\begin{align}
c_k = \left(
\frac{10}{9\pi}, 
\frac{44-8\ln 2}{21\pi^2},
0.0303089(0),
-0.0708(1),\ldots
\right),
\label{eq:cns}
\end{align}
where the ``(0)'' after the seventh digit of the third-order coefficient means that its numerical error is smaller than $5\cdot 10^{-8}$.
Notably, for spins higher than $1/2$ both
logarithmic terms $\sim x^{n\geqslant 4}\ln|x|$ 
and multi-fermion couplings are present in the perturbation series~\cite{Wellenhofer:2018dwh}.
The extrapolation problem is then much more intricate in that case, 
in particular since more and more multi-fermion parameters appear at higher orders
(the complete perturbation series has an infinite number of them).
Therefore, in this paper we consider only the spin-$1/2$ case with the single parameter $a_s$. The
perturbation series is then analytic in $x=\kF a_s$.

The strong-coupling regime of the dilute Fermi gas has been the focus of many experimental and theoretical studies in the past 
two decades~\cite{RevModPhys.80.885,RevModPhys.82.1225,BCSBECtheory}. 
Here, the salient feature at zero temperature is 
the BCS-BEC crossover from large negative to large positive $a_s$, 
where the point with infinite scattering length (i.e., $1/a_s=0$) 
is referred to as the unitary limit.
By dimensional analysis, the energy of the unitary Fermi gas is given by $E(\kF)=\xi E_0(\kF)$, where $\xi$ is called the Bertsch parameter.

Regarding the CEP, what is relevant is not the functional form of the $\kF$ 
dependence at $1/a_s=0$ but only that $\xi$ is the leading term in the strong-coupling expansion [Eq.~\eqref{eq:sce}].
The Bertsch parameter has been
determined experimentally with ultracold atoms as 
${\xi=0.376(4)}$~\cite{Ku563} and from quantum Monte Carlo (QMC) simulations as
${\xi=0.372(5)}$~\cite{PhysRevA.84.061602}.
Further, QMC provides estimates for the leading two strong-coupling coefficients $d_1$ and $d_2$, $d_1 \approx -0.9$ and $d_2 \approx -0.8$,
with $d_1$ known more precisely~\cite{PhysRevA.83.041601} (see also Ref.~\cite{Navon729}). 
Such a situation is typical of many physics problems since often only limited data are available in the nonperturbative region.

The present paper is organized as follows.
First, in Sec.~\ref{sec:Pade} we apply (two-point) Pad\'e approximants
to the CEP for the dilute Fermi gas. 
We find that Pad\'es do not yield satisfactory results for the CEP, 
in particular several of them give flawed approximants with poles in the BCS region.
In Sec.~\ref{sec:Borel} we then examine Borel methods.
We find these methods also have severe deficiencies regarding our goal of producing well converged classes of constrained extrapolants.
Next, in Sec.~\ref{sec:MaxEnt} we study Maximum Entropy extrapolation (MaxEnt)~\cite{BenderMeadPapa,Drabold_1991}, which
is a modification of Borel extrapolation that allows the definition of a larger class of approximant functions.
It has been demonstrated that in certain cases MaxEnt 
can outperform both Borel and Pad\'e methods~\cite{BenderMeadPapa,Drabold_1991}, and we examine one such case, the harmonic oscillator with an octic term. 
However, we show that to be applicable to the dilute Fermi gas MaxEnt  requires extensive elaborations that render it impractical in that case.

After these unsuccessful explorations, in Sec.~\ref{sec:ODME} we then
introduce order-dependent-mapping extrapolation (ODME) as a straightforward and very flexible 
method for the CEP.
The ODME method, which was developed in Ref.~\cite{Wellenhofer:2020ylh},
improves on the order-dependent mapping (ODM) approach invented
by Seznec and Zinn-Justin~\cite{Seznec:1979ev} (see also Ref.~\cite{Yukalov:2019nhu}) and builds
in information on the leading strong-coupling coefficients---$d_1$ and $d_2$ in this case. We review the connection of the original ODM to optimized perturbation theory and discuss the principles used in previous work to select the parameter $\alpha$ of the mapping. We then explain why, in the CEP, strong-coupling information provides a physics-guided way to choose $\alpha$ and show that ODME leads to well converged approximants for
the ground-state energy of the dilute Fermi gas. The ODME solution to the CEP for the dilute Fermi gas compares
well with QMC results throughout the BCS regime and beyond.
Finally, Sec.~\ref{sec:conclusion} offers a summary and avenues for future work. 

Some of these results already appeared in Ref.~\cite{Wellenhofer:2020ylh}, but here we provide additional discussions of them, as well as detailed comparisons to the Borel and MaxEnt methods.

\section{Pad\'e approximants}

\epigraph{\textit{``It's a dangerous business, Frodo, extrapolation. You step out of the perturbative regime, and if you don't keep your analytic structure, there's no knowing where you might be swept off to.”}}

\label{sec:Pade}

Pad\'e extrapolation works by fixing the coefficients of a rational function $\text{Pad\'e[$n$,$m$]}(x)$ 
such that its Maclaurin series matches the perturbation series to order $N=n+m$.
Here, $n$ is the degree of the polynomial in the numerator, 
$m$ is the denominator degree, 
and the Pad\'e approximant is normalized to $\text{Pad\'e[$n$,$m$]}(0)=1$ in our case.
Two-point Pad\'es are the generalization of this approach to the CEP:
the Pad\'e coefficients are matched to both the perturbation series and the strong-coupling expansion up to specified orders $N$ and $M$. According to the definition of the CEP in the introduction $M$ remains fixed for each $N$. $N$ and $M$ are related to the degrees of the polynomials in the Pad\'e according to
$N+M=n+m-2$.
As the Bertsch parameter is finite and nonzero, 
we are restricted to
``diagonal Pad\'es" with $n=m$, i.e., 
\begin{align}  \label{Padeform}
\text{Pad\'e[$n$,$m=n$]}(x) = \frac{1+ \sum_{k=1}^n a_k x^k}{1+\sum_{k=1}^n b_k x^k} \,.
\end{align}
Since for diagonal Pad\'es $N+M=2n-2$ is even, a further impediment is that 
they are applicable to the CEP only for even or odd truncation orders, respectively (depending on $M$). 
Moreover, a general problem with Pad\'e approximants is that they can have spurious poles in the region of interest. 
A simple example is given in the book by Bender and Orszag~\cite{benderbook}: the $\text{Pad\'e[$N$,$1$]}$ approximant for the function 
$y(x)=(x+10)/(1-x^2)\stackrel{x\rightarrow 0}{\simeq} \sum_{n=0}^N {a_n x^n}$ is given by
\begin{align}  \label{Padeform2}
\text{Pad\'e[$N$,$1$]}(x) = \sum_{n=0}^{N-2} {a_n x^n}+\frac{a_{N-1}x^{N-1}}{1-a_Nx/a_{N-1}} \,.
\end{align}
This has a simple pole at $x=1/10$ if $N$ is even, and this feature is absent in the exact $y(x)$.

\begin{figure*}[t] 
\centering
\includegraphics[width=0.91\textwidth]{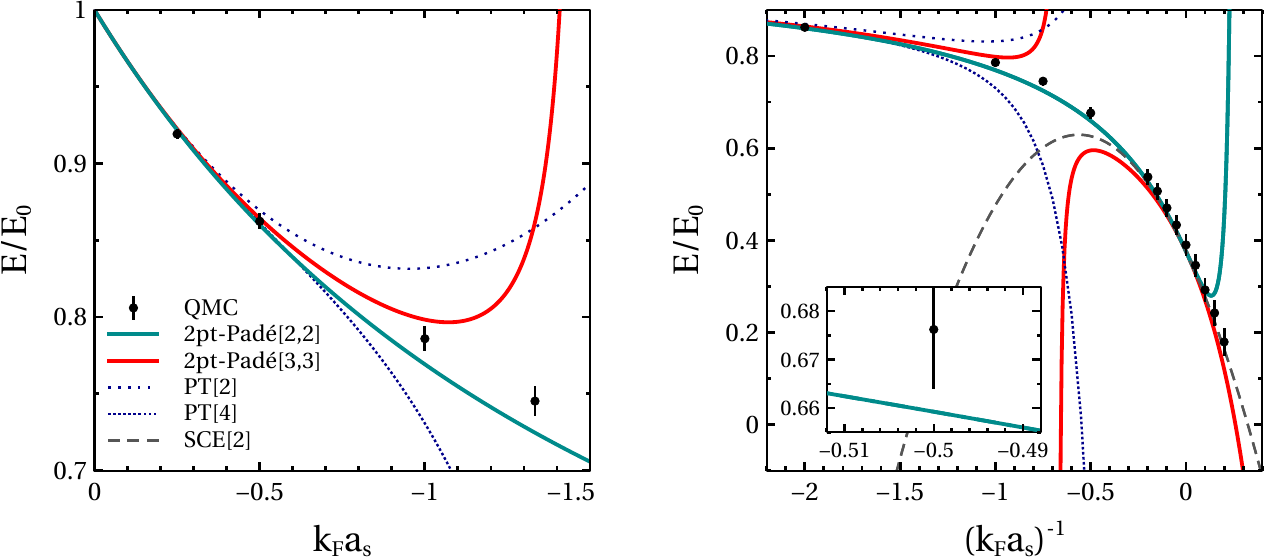} 
\caption{Two-point Pad\'e approximants for the ground-state energy $F(x)$ of the dilute Fermi gas, see text for details. 
The left plot shows the weak-to-intermediate coupling BCS regime of $F(x)$ as a function of $x=\kF a_s$, while 
the right plot depicts $F(x)$ as a function of $1/x=(\kF a_s)^{-1}$ throughout the BCS-BEC crossover. 
In each plot we also show the truncated perturbation theory (PT) series PT$[N]$ for ${N=2,4}$, 
and in the right plot also the strong-coupling expansion (SCE) truncated at second order ($M=2$).
In both cases we also show results from QMC computations~\cite{Gandolfi:2015jma}. 
(The errors of the QMC data are estimated as specified in Ref.~\cite{Wellenhofer:2020ylh}.)
The inset in the right plot magnifies the behavior at the intermediate coupling $x=-2$.}
\label{fig:Padeplot}
\end{figure*}

In Fig.~\ref{fig:Padeplot} we show the results for the ground-state energy $F(x)$ of the dilute Fermi gas obtained from diagonal Pad\'e approximants 
matched to $\xi$, $d_1$ and $d_2$, and $N=2,4$ perturbation coefficients.
In addition, we also show results obtained from the truncated perturbation series and the truncated strong-coupling expansion, as well as results from QMC computations~\cite{Gandolfi:2015jma}.
One sees that the perturbation series provides well converged results for $|x|\lesssim 0.5$~\cite{Wellenhofer:2018dwh} and diverges strongly for $|x|\gtrsim 1$.
The behavior of the strong-coupling expansion is similar, i.e., it is well-converged for $|1/x|\lesssim 0.5$ only.
Regarding the $N=2,4$ two-point Pad\'es, ``$\text{2pt-Pad\'e[$2$,$2$]}$'' and ``$\text{2pt-Pad\'e[$3$,$3$]}$'', 
at weak coupling they both improve upon the perturbative approximants: 
for $|x|\lesssim 1$ they are close to each other and near the QMC results, with the higher-order Pad\'e more accurate.
The low-order ($N=2$) Pad\'e actually provides reasonably good results throughout the BCS region, i.e., it is not too far off the QMC data for $1/x < 0$.
Eventually though, there is 
a spurious pole at $1/x=0.27$, so in 
the BEC region it fails to provide viable results already for small positive $1/x$.
The $N=4$ Pad\'e on the other hand
has a pole 
at negative real coupling (at $x\approx -1.49$), 
and as can be seen in Fig.~\ref{fig:Padeplot} this leads to a significantly impaired extrapolation.
We note that the $n=3$ two-point Pad\'e constructed to match $\xi$, $d_1$, $d_2$, and three perturbation coefficients also has a pole at negative real coupling, at $x\approx-0.17$.
Because of these defects in the analytic structure we conclude that two-point Pad\'e approximants are not suitable for the CEP of the dilute Fermi gas.

\section{Borel extrapolation methods}
\epigraph{\textit{``Long years ago, in the last years of the nineteenth century,  Mathematicians forged resummation methods of great power.''}}

\label{sec:Borel}

\subsection{General discussion}

Compared to Pad\'e approximants,
Borel extrapolation represents a more sophisticated framework that allows the construction of approximants that explicitly take into account 
the large-order behavior of the perturbation series.
The starting point of Borel methods is the Borel(-Leroy) transformed perturbation series
\begin{align} \label{Borelseries}
\mathcal{B}(t)
\stackrel{t\rightarrow 0}{\simeq}
1+\sum_{k=1}^\infty \frac{c_k}{\Gamma(k+1+\beta_0)}t^k\,,
\end{align}
where the standard Borel transform corresponds to $\beta_0=0$.
In contrast to the perturbation series, the Borel transformed series has a finite convergence radius.
That is, the
large-order behavior 
\begin{align} \label{PTlargeorder}
c_k\stackrel{k\rightarrow \infty}{\sim} a^k\Gamma(k+1+\beta)\,,
\end{align} 
together with the choice of $\beta_0$,
determines the nature of the leading singularity of $\mathcal{B}(t)$
at ${t=1/a}$, see Refs.~\cite{Kleinert:2001ax,Costin_2019,PhysRevD.100.056019}.

Borel resummation corresponds to constructing $F(x)$
in terms of the inverse Borel transform $B(x)$ of
the analytic continuation $\mathcal{B}(t)$ of the Borel transformed perturbation series:
\begin{align}\label{Laplacetraf}
B(x) =\int_{0}^\infty  \!\! dt \e^{-t} t^{\,\beta_0} \mathcal{B}(tx)\,. 
\end{align}
In the so-called ``Borel non-summable case'' where $\mathcal{B}(tx)$ has poles on the positive real axis one can 
shift the integration path infinitesimally off the real axis.
In this case the approximant for $F(x)$
may be taken as the real part of $B(x)$, a prescription that often gives the correct result~\cite{ZinnJustin:1980uk,Seznec:1979ev,doi:10.1063/1.4921155,Marino2019b}.
The detailed study of the analytic properties of 
the Borel transform and the Borel non-summable case are the subject of resurgence theory~\cite{DORIGONI2019167914,ANICETO20191}.

In the realistic case where the perturbation series is known only to a certain order $N$, 
the goal is to construct approximants $B_N(x)$ for $F(x)$ 
via the inverse Borel transform of approximants $\mathcal{B}_N(t)$ for $\mathcal{B}(t)$.
There are three basic methods for that:\footnote{
As discussed, in each of these three methods we incorporate the unitary limit $F(-\infty)= \xi$.
In the CB and PCB case, this corresponds to the 
methods we denoted as CCB and PCCB in Ref.~\cite{Wellenhofer:2020ylh},
where the additional ``C'' stands for ``constrained''.
We note that in the weak-to-intermediate coupling region (i.e., for $|x|\lesssim 1$), 
the ``unconstrained'' Borel methods give results that are similar to the ones obtained from the ``constrained'' Borel methods.}
\begin{itemize}
\item Pad\'e-Borel: matching of Pad\'e approximants to the Borel transform of the truncated perturbation series, 
\item conformal-Borel (CB): truncated re-expansion of the Borel transform series in terms of a conformal 
mapping $w(t)$ that is chosen based on the analytic properties of $\mathcal{B}(t)$,
\item Pad\'e-conformal-Borel (PCB): matching of Pad\'e approximants to the conformally re-expanded Borel series.
\end{itemize}
For each of these, to incorporate the correct unitary limit $B_N(-\infty)= \xi$ we introduce a
rescaled version of $F(x)$
that approaches $0$ as $x \rightarrow -\infty$:
\begin{equation} \label{fscale}
f(x)=\frac{F(x)-\xi}{1-\xi} \,.
\end{equation}
The implementation of further strong-coupling constraints is less straightfoward, and not considered here. 
A study of this problem can be found in Ref.~\cite{HONDA2015533}, where it was found that two-point Pad\'{e}-Borel approximants do not 
improve upon two-point Pad\'{e}s.

In Pad\'e-Borel, incorporating the unitary limit $f(-\infty)=0$ requires that off-diagonal $\text{Pad\'e[$n$,$m$]}$ approximants with $m>n$ are used.
Pad\'e-Borel, in contrast to the conformal methods CB and PCB, makes no use of analytic properties of $f(x)$.
In CB and PCB the principal analytic information determining the choice of the 
conformal mapping $w(t)$ is the large-order behavior of the perturbation series, specifically the coefficient $a$ in Eq.~\eqref{PTlargeorder}.
That is, given knowledge of $a$ one chooses the mapping
\begin{align} \label{cBmap}
w(t)=\frac{\sqrt{1-a t}-1}{\sqrt{1-a t}+1}
\end{align}
that maps the cut Borel $t$-plane to the interior of the unit disc~\cite{ZinnJustin:1980uk,PhysRevD.100.056019}.
The conformal reexpansion of the Borel series is then constructed as
\begin{align} \label{CB}
\mathcal{B}_N(t)=\bigl(1-w(t)\bigr)^\eta \sum_{k=0}^N s_k [w(t)]^k\,,
\end{align} 
where the prefactor ensures the correct unitary limit $B_N(-\infty)=0$. The exponent $\eta$ of the prefactor in Eq.~\eqref{CB} is chosen such that the known analytic structure at infinity is best reproduced.
Since the strong-coupling expansion [Eq.~\eqref{eq:sce}] has no fractional powers of $1/x$, we set $\eta=1$. 
The coefficients $s_k$ are given by
\begin{align} \label{eq:s_k}
s_k=\frac{1}{(1-\xi)k!}\sum_{n,m=0}^{k} c_n  \gamma_{n,m}(0) \,,
\end{align}
with
\begin{align}
\gamma_{n,m}(t)=  \frac{\partial^m [t(w)]^n}{\partial w^m}\bigg|_{w=w(t)}\,.
\end{align} 
The conformal transformation~\eqref{cBmap} 
yields a function that has a square-root branch point at ${t=1/a}$.
A refinement of conformal Borel resummation (compared to the standard choice $\beta_0=0$)
corresponds to setting ${\beta_0=\beta+3/2}$, since then   
the exact Borel transform has the same feature~\cite{Kleinert:2001ax}.

\begin{figure*}[p] 
\centering
\includegraphics[width=0.85\textwidth]{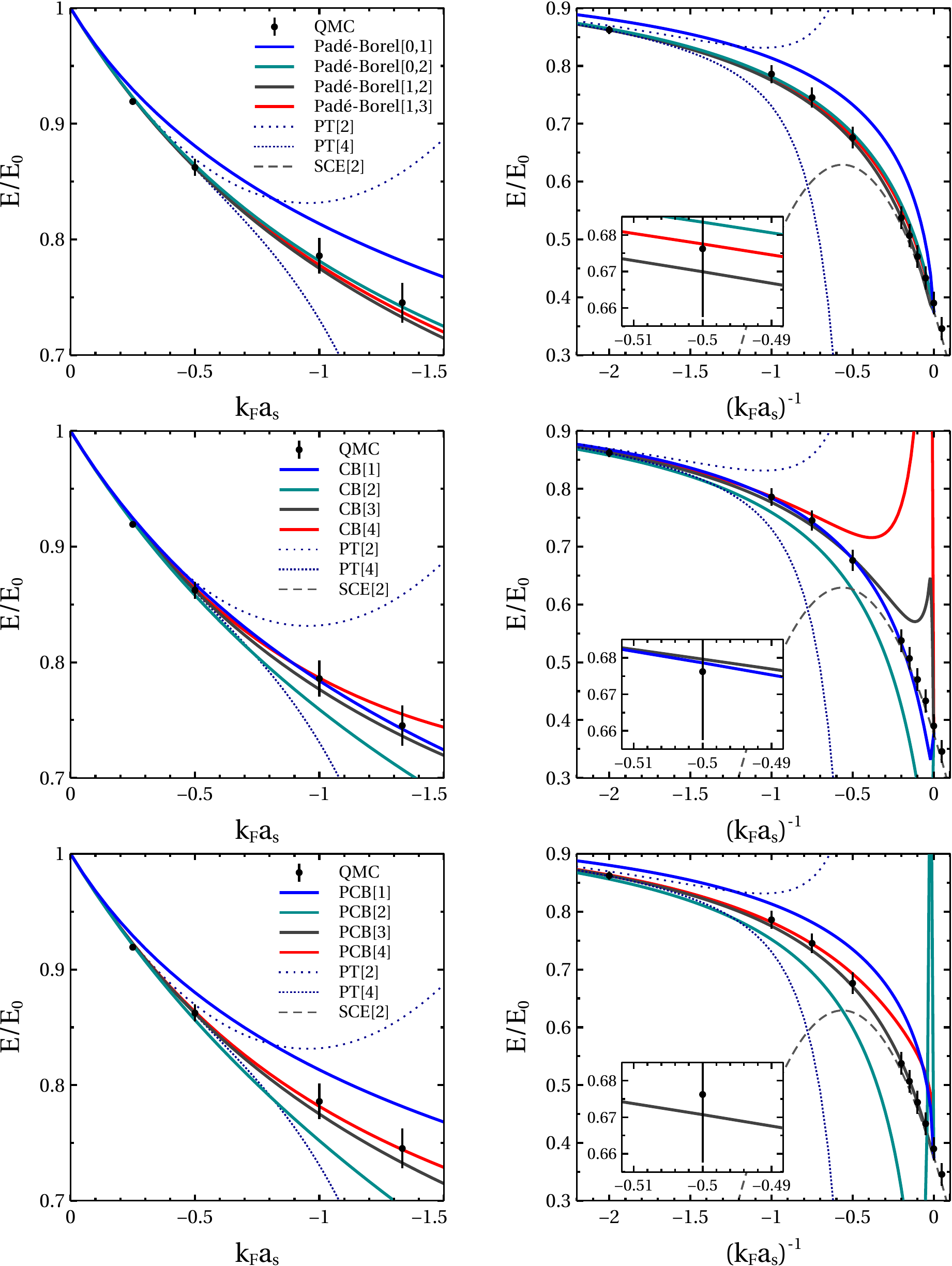} 
\caption{Different Borel approximants for the ground-state energy $F(x)$ of the dilute Fermi gas, see text for details. 
As in Fig.~\ref{fig:Padeplot}, the two plots in each row depict different parts of the BCS regime. 
In the strong-coupling plot (right-hand side) we cut out most of the BEC region since Borel approximants 
cannot easily be continued beyond the unitary limit.
The first row shows various Pad\'e-Borel approximants with $\beta_0=0$, while the subsequent two rows display results obtained from the CB and PCB methods with $\beta_0=3/2$.
As in Fig.~\ref{fig:Padeplot}, we also show perturbation theory (PT), the strong-coupling expansion (SCE), and the QMC results.}
\label{fig:Borelplot}
\end{figure*}

\subsection{Application to dilute Fermi gas}

The large-order behavior of the perturbation series of the dilute Fermi gas is not known at present.
The basic idea is that the (leading) large-order behavior should be determined by (dominant) nonperturbative properties of the system---that would be pairing effects in this case of the dilute Fermi gas.
Based on their results for the dilute Fermi gas in 1D~\cite{Marino2019,Marino2019b}, 
Mari{\~{n}}o and Reis recently conjectured that ${a=-1/\pi}$ and ${\beta=0}$ for the 3D Fermi gas~\cite{Marino2019b}.
This conjecture is also motivated by their finding that the subset of ladder diagrams gives ${a=-1/(2\pi)}$, while
the subset of 
particle-particle ladders yields ${a=-1/(4\pi)}$~\cite{Marino2019b}.

The results from different Borel approximants for the ground-state energy $F(x)$ of the dilute Fermi gas 
are depicted in Fig.~\ref{fig:Borelplot}.
Regarding the two choices for the 
parameter $\beta_0$ in the Borel transform [Eq.~\eqref{Borelseries}] considered above, 
for Pad\'e-Borel we use the standard choice $\beta_0=0$ 
while for CB and PCB we use the refined choice $\beta_0=\beta+3/2=3/2$.
In each case, this yields results that are closer to the central QMC points.
(Note that the refined choice is motivated only for CB and PCB but not for Pad\'e-Borel.)
Further, in contrast to the case of Pad\'e approximants, no simple analytic continuation into the BEC region 
is available for Borel approximants.

Regarding Pad\'e-Borel$[n,m]$ approximants matched to the rescaled function $f(x)$, in the upper row of Fig.~\ref{fig:Borelplot}
we show the results obtained for $[n,m]\in\{[0,1],[0,2],[1,2],[1,3]\}$, corresponding to 
input from the first $N\in\{1,2,3,4\}$ perturbative coefficients.
One sees that at weak-to-intermediate coupling, 
the results from the $N=2,3,4$ Pad\'e-Borels are within the QMC errors, 
The $N=3,4$ results are within the QMC errors throughout the entire BCS regime.
For $|x|\lesssim 1.5$  the $N=2$ results are closest to the central QMC points, 
while for larger $|x|$ the $N=4$ ones perform better.

The results from the CB$[N]$ and PCB$[N]$ approximants,\footnote{In the PCB case we use the standard $[n,n]$ and $[n,n+1]$ Pad\'es for even and odd $N$, respectively.} shown in the middle and lower row of Fig.~\ref{fig:Borelplot}, are notably different. 
At low $|x|\lesssim 1.5$, the CB results that come closest to the central QMC points are those obtained for $N=1$, followed by $N=3$ and $N=4$.
In the strong-coupling regime all CB approximants deviate substantially from the QMC data at some point. 
Since the correct unitary limit is enforced, this behavior is accompanied by local extrema. 
In the PCB case, for $|x|\lesssim 1.5$ the results improve with increasing $N$, with the $N=3,4$ ones consistent with the QMC data.
However, at larger $|x|$ only the $N=3$ PCB stays within the QMC error bars.

Altogether, we find that among the Borel methods only Pad\'e-Borel approximants give good results in the BCS region.
However, no simple analytic continuation into the BEC region and no straightforward way to incorporate strong-coupling information beyond the unitary limit is available for them.
Further, we note that
although Pad\'e-Borel approximants perform reasonably well for the dilute Fermi gas, in other cases such
as the 0D model and the 1D Fermi gas considered in Ref.~\cite{Wellenhofer:2020ylh} they do not give 
accurate results at strong coupling (see Figs.~A.1 and A.2 of Ref.~\cite{Wellenhofer:2020ylh}).\footnote{To be 
precise, in Ref.~\cite{Wellenhofer:2020ylh} we used the standard $[n,n+1]$ (for odd $N$) and $[n,n]$ (for even $N$) Pad\'es for Pad\'e-Borel, 
whereas here we use $[n,n+1]$ (for odd $N$) and $[n,n+2]$ Pad\'es.
For the 0D and 1D cases considered in Ref.~\cite{Wellenhofer:2020ylh}, the $[n,n+2]$ Pad\'e-Borel approximants 
improve upon the $[n,n]$ ones only near the strong-coupling limit.}
Therefore, we conclude that Borel methods are only of limited suitability for this CEP.

\section{Maximum Entropy extrapolation (MaxEnt)}
\epigraph{\textit{``Fermi gas drowns out all but the brightest extrapolations.''}}

\label{sec:MaxEnt}

\subsection{Setup}

Maximum Entropy extrapolation (MaxEnt)~\cite{BenderMeadPapa,Drabold_1991,Drabold1991} may be seen as a generalization of Borel extrapolation. 
That is, instead of Eq.~\eqref{Laplacetraf} we consider approximants $f_N(x)$ 
for $f(x)$ of Eq.~\eqref{fscale}
of the more general form 
\begin{align} \label{eq:MEansatz}
f_N(x)= \int_D dt \, \rho(t) \, K(t x)\,,
\end{align}
where the density $\rho(z)$ is semipositive in the domain $D$, i.e., $\rho(z)\geqslant 0$ for $z\in D$.
Since we consider the rescaled function $f(x)$ given by Eq.~\eqref{fscale}, 
the kernel $K(x)$ should satisfy $K(-\infty)=0$.
Given a choice of the kernel $K(x)$ and the domain $D$, 
the approximant $f_N(x)$ is then required to reproduce the first $N$ terms of the perturbation series of $F(x)$.
This means that one needs to solve the (finite) moment problem
\begin{align} \label{eq:MEmoments}
\forall n\in\{0,\ldots,N\}:\;\;\int_D dt \, \rho(t) \, t^n = \mu_n\,,
\end{align}
where the moments are given by
$\mu_n=c_n/K_n$, with $K_n$ the Maclaurin coefficients of $K(x)$. 
Without loss of generality we may choose a function $K(x)$ that has $K_0=1$, so since $F(x)$ (or $f(x)$) have $c_0=1$ we may assume $\mu_0=1$.

The moment problem [Eq.~\eqref{eq:MEmoments}] is solvable only in certain cases, as discussed below.
However, even if it is solvable, the moment conditions do not determine the density $\rho(t)$ uniquely.
The MaxEnt idea is to pick the most likely solution for $\rho(t)$.
This means that as an additional condition on $\rho(t)$
we require that the entropy functional $S[\rho]$ given by
\begin{equation} \label{eq:MEentropy}
S[\rho]=-\int_D dt \, \rho(t) \ln(\rho(t))
\end{equation}
is maximized~\cite{PhysRev.106.620,PhysRev.108.171}.
Via the method of Lagrange multipliers, the moment problem and $\delta S/\delta\rho=0$ can be combined into
the condition
\begin{equation} \label{eq:MEcondition}
\frac{\delta S_N[\rho]}{\delta\rho}=0\,,
\end{equation}
with
\begin{equation} \label{eq:MESN}
S_N[\rho]=S[\rho] + \sum_{n=0}^N \lambda_n \left(\int_D dt \, \rho(t) \, t^n - \mu_n\right)\,.
\end{equation}
The (formal) solution of Eq.~\eqref{eq:MEcondition} is readily found as
\begin{equation} \label{eq:MErhoN}
\rho_N(t)=\exp\left(-\sum_{n=0}^N \lambda_n t^n\right)\,,
\end{equation}
where the Lagrange multipliers $\lambda_n$ are determined by the solution of the moment problem
with $\rho_N(t)$ substituted for $\rho(t)$, i.e.,
\begin{align} \label{eq:MEmaxentmoments}
\forall n\in\{0,\ldots,N\}:\;\;\int_D dt \, \exp\left(-\sum_{m=0}^N \lambda_m t^m\right) \, t^n = \mu_n\,.
\end{align}
The problem of solving these $N+1$ nonlinear equations can be formulated as a minimization problem~\cite{MeadPapa}.
That is, we define the ``effective potential'' $\Gamma(\lambda_1,\ldots,\lambda_N)$ as
\begin{equation}
\Gamma=\ln Z  + \sum_{n=1}^N \mu_n \lambda_n\,,
\end{equation}
where
\begin{equation}
Z=\int_D dt \, \e^{-\sum_{n=1}^N \lambda_n t^n}\,.
\end{equation}
From Eq.~\eqref{eq:MEmaxentmoments}
together with $\mu_0=1$ it follows that $Z = e^{\lambda_0}$, i.e., 
$\lambda_0$ is determined by $\{\lambda_1, \ldots, \lambda_N\}$.
Stationary points of $\Gamma$ obey the moment conditions, and $\Gamma$ can be shown to be convex everywhere~\cite{MeadPapa}. 
Thus, given that $\Gamma$ has a local minimum, 
this unique minimum yields the set $\{\lambda_1, \ldots, \lambda_N\}$ that solves the MaxEnt moment problem, Eq.~\eqref{eq:MEmaxentmoments}.

\subsection{Moment conditions}

Here, we discuss the necessary and sufficient conditions for the existence of a (unique) solution to the MaxEnt moment problem.
We consider a compact domain $D=[a,b]$, so Eq.~\eqref{eq:MEmoments} is the finite (i.e., $N$ is finite) Hausdorff moment problem~\cite{akhiezerbook}.
We refer to Eq.~\eqref{eq:MEmaxentmoments} as the (finite) Hausdorff MaxEnt moment problem.
The case where  $D=[a,\infty]$ is known as the Stieltjes moment problem (and $D=[-\infty,\infty]$ corresponds to the Hamburger moment problem).
Notably, the Stieltjes moment conditions~\cite{DowsonWragg,Tagliani} deviate from the $b\rightarrow \infty$ limit of the Hausdorff conditions discussed below.
However, since in the Stieltjes case it is $\rho(t)\rightarrow 0$ for $t\rightarrow \infty$, for practical applications 
we can assume a compact domain.

The Hausdorff moment problem can be rescaled to the canonical form
\begin{align} 
	\forall n\in\{0,\ldots,N\}:\;\;\int_{0}^{1} \!\! d\zeta \, p(\zeta) \,\zeta^n = \omega_n\,.
\end{align} 
Here, the scaled moments $\omega_n$ are obtained iteratively from 
the $N+1$ linear equations
\begin{align} 
   \forall n\in\{0,\ldots,N\}:\;\;\sum_{k=0}^n  \binom{n}{k} (b-a)^k \omega_k a^{n-k}=\mu_n\,.
\end{align} 
 The original density $\rho(t)$ is obtained from the scaled density $p(\zeta)$ using the affine transformation that relates $t$ and $\zeta$:
\begin{align} 
\rho(t)=\frac{1}{b-a} \, p\left(\frac{t-a}{b-a}\right)\,.
\end{align} 

As shown in the book by Akhiezer (p.~74 of Ref.~\cite{akhiezerbook}),
the necessary and sufficient condition for the existence of a unique solution to the finite Hausdorff moment problem is that the eigenvalues of the matrices
\begin{align} 
Q=\big(\omega_{i+j+1}\big)_{i=0,\ldots,m,\,j=0,\ldots,m},\quad\quad
R=\big(\omega_{i+j}-\omega_{i+j+1}\big)_{i=0,\ldots,m,\,j=0,\ldots,m},
\end{align} 
for  odd $N=2m+1$ and 
\begin{align} 
Q=\big(\omega_{i+j}\big)_{i=0,\ldots,m,\,j=0,\ldots,m},\quad\quad
R=\big(\omega_{i+j+1}-\omega_{i+j+2}\big)_{i=0,\ldots,m-1,\,j=0,\ldots,m-1},
\end{align} 
for even $N=2m$
are non-negative.
For $N\in\{1,2,3\}$, this (together with $0\leqslant \omega_{n+1}\leqslant\omega_n\leqslant 1$) yields the following conditions on the scaled moments $\omega_n$:
\begin{align} \label{MEcondhausdorff1}
N=1:\;\;\;\; & 0\leqslant \omega_1\leqslant 1\,, \\ \label{MEcondhausdorff2}
N=2:\;\;\;\; & 0\leqslant \omega_1^2\leqslant \omega_2\leqslant \omega_1\leqslant 1\,, \\ \label{MEcondhausdorff3}
N=3:\;\;\;\; & 
0\leqslant \frac{\omega_2^2}{\omega_1}\leqslant \omega_3\leqslant \omega_2\leqslant \omega_1\leqslant 1 \;\; 
\land \;\; 
\omega_1^2-\omega_1 (\omega_2+\omega_3)-\omega_2 (1-\omega_2)+\omega_3 \leqslant 0\,.
\end{align}
In Ref.~\cite{MeadPapa}, Mead and Papanicolaou
proved that the
existence conditions for the infinite Hausdorff moment problem (i.e., complete monotonicity of the sequence 
$\{\omega_n\}$, $n\in\{1,\ldots,\infty\}$) 
guarantee the existence of a unique solution of the associated infinite Hausdorff MaxEnt moment problem. 
Their proof can be trivially adapted to prove that the existence conditions for the finite Hausdorff moment problem 
are sufficient also for the existence of a unique solution of the 
finite Hausdorff MaxEnt problem.

\subsection{Application to octic oscillator}

Before we investigate 
the requirements on the kernel $K(x)$ and the domain $D=[a,b]$ imposed by the Hausdorff moment conditions for the dilute Fermi gas, 
following Refs.~\cite{BenderMeadPapa,Drabold_1991} we first study a different system where these conditions are easily satisfied: the harmonic oscillator with an octic term, which we refer to as the octic oscillator.

The octic oscillator is defined by the Hamiltonian
\begin{align}  
H=p^2/2+x^2/2+g x^8\,,
\end{align}
where $[x,p]=i$.
Following  Bender \emph{et al.}~\cite{BenderMeadPapa} we write the ground-state energy as $E(g)=1/2+105 g f(g)/16$, where
$f(g)\rightarrow 0$ and $E(g)\rightarrow\infty$ for $g\rightarrow \infty$.
The weak-coupling perturbation series of $f(g)$ reads
\begin{align}  
f(g)\stackrel{g\rightarrow 0}{\simeq}
1+\sum_{k=1}^N c_k g^k + O(x^{N+1}) \,.
\end{align}
The leading perturbation coefficients are given by
\begin{align}
c_k= \left(
-\frac{643}{2},\,
\frac{3 824 275}{8},\,
-\frac{242 134 255 883}{128},\,
\frac{8 050 560 668 350 165}{512},\,
-\frac{973 733 659 602 733 224 723}{4096},
\ldots 
\right)
\end{align}
and their large-order behavior is of the form~\cite{BenderMeadPapa}
\begin{align}  
c_k \stackrel{k\rightarrow \infty}{\sim} k^{-1/2} q^k (3k)!\,.
\end{align}
It can be shown that $f(g)$ is a generalized Stieljes function~\cite{BenderMeadPapa}. That is, $f(g)$
can be written as
\begin{align}  \label{octicStielt}
f(g)=
\int_0^\infty \!\! dt\,\frac{\rho(t)}{1+t g} \,,
\end{align}
so the perturbation coefficients can be represented as 
\begin{align}   \label{octicStieltmom}
c_k = \int_0^\infty \!\! dt \, \rho(t) t^{k-1}.
\end{align}
Because of the rapid growth of the perturbation coefficients with $k$, the density $\rho(t)$ is not uniquely 
determined~\cite{BenderMeadPapa}.
However, Eq.~\eqref{octicStieltmom}
implies that the (Stieltjes) MaxEnt moment problem
with kernel $K_0(g)=1/(1+g)$ does have a unique solution (for each $N$).
Thus, we consider MaxEnt approximants $f_N(g)$ for $f(g)$
of the form
\begin{align}  \label{octicMaxEnt}
f_N(g)=
\int_0^\infty \!\! dt\, \e^{-\sum_{n=0}^N \lambda_n t^n} K_0(tg) \,,
\end{align}
where the kernel 
\begin{align} \label{Kernel1}
K_0(g)=\frac{1}{1+g}
\end{align}
associated with Eq.~\eqref{octicStielt}. This
corresponds to the original kernel choice by Bender \emph{et al.}
in their pioneering study of the MaxEnt extrapolation technique~\cite{BenderMeadPapa}.
An improved kernel that incorporates the leading strong-coupling behavior~\cite{doi:10.1063/1.523061}
\begin{align} 
f(g)\stackrel{g\rightarrow \infty}{\sim} g^{-4/5}
\end{align}
was considered by Drabold and Jones~\cite{Drabold_1991}, i.e., 
\begin{align}  \label{Kernel2}
K_\text{DJ}(g)=\frac{1}{(1+g)^{4/5}}\,.
\end{align}
In Table~\ref{tableoctic}
we compared the low-order MaxEnt results obtained from these two kernel choices
to exact results.
Also shown are results from Pad\'e$[n,n+1]$ approximants. Note that although $E(g)$ is a Stieltjes function, 
since $c_k$ grows more rapidly than $(2k)!$
the Pad\'e approximants are not guaranteed to converge to the correct answer~\cite{BenderMeadPapa}.
One sees that the improved kernel of Drabold and Jones $K_\text{DJ}(g)$ gives better results than the kernel of Bender \emph{et al.}, $K_0(g)$. For both kernels, MaxEnt performs significantly better than Pad\'e approximants.

\begin{table}
\caption
{Results from MaxEnt and Pad\'e approximants for the inverse of the octic oscillator function $f(g)$ are compared to the exact $1/f(g)$ for 
different perturbation orders $N$ 
and ${g\in\{0.001,0.01,0.1,0.5,1,10\}}$. The exact results are taken from Refs.~\cite{BenderMeadPapa,doi:10.1063/1.523061,Gayathri}.}
\begin{center}
\begin{ruledtabular}
\begin{tabular}{l l ll lllll}
\addlinespace
method 			   		   		   &$N$	    && $g=1/1000$ 	& $g=1/100$	    & $g=1/10$ 		& $g=1/2$ 	& $g=1$ 	& $g=10$ 			\\
\hline
exact			  				   &---		&& 1.21			& 2.04	    	& 5.46			& 13.37		& 20.46 	& 94.98				\\
\hline
MaxEnt[$K_\text{DJ}$]			   &$1$		&& 1.25			& 2.49			& 8.47			& 24.20		& 39.18 	& 209.9 			\\
MaxEnt[$K_\text{DJ}$]			   &$2$		&& 1.24			& 2.41			& 8.02			& 22.43		& 36.75		& 195.9				\\
MaxEnt[$K_\text{DJ}$]			   &$3$		&& 1.22			& 2.25			& 7.14			& 19.83		& 31.83		& 167.6				\\
\hline
MaxEnt[$K_0$]					   &$1$		&& 1.26			& 2.68			& 10.66			& 35.43		& 61.65		& 428.6				\\
MaxEnt[$K_0$]					   &$2$		&& 1.25			& 2.57			& 9.94			& 32.68		& 56.68		& 391.1				\\
MaxEnt[$K_0$]					   &$3$		&& 1.22			& 2.37			& 8.61			& 27.58		& 47.44		& 321.8				\\
\hline
Pad\'e							   &$1$		&& 1.32			& 4.22			& 33.16			& 161.9		& 322.5		& 3216				\\
Pad\'e							   &$3$		&& 1.24			& 3.37			& 24.49			& 118.4		& 235.7		& 2348				\\
Pad\'e							   &$5$		&& 1.23			& 3.08			& 21.49			& 103.3		& 205.5		& 2046				\\
Pad\'e							   &$7$		&& 1.23			& 2.94			& 19.92			& 95.4		& 189.8		& 1888				\\
Pad\'e							   &$9$		&& 1.22			& 2.85			& 18.95			& 90.5		& 179.9		& 1789				\\
\end{tabular}
\end{ruledtabular}
\end{center}
\label{tableoctic}
\end{table}

\subsection{Impracticality for dilute Fermi gas}


We now consider the application of the MaxEnt extrapolation technique for the CEP of the dilute Fermi gas, i.e., 
for the rescapled function $f(x)$ given by Eq.~\eqref{fscale}.
In terms of the unscaled moments, the ${N\in\{1,2\}}$ Hausdorff conditions [Eqs.~\eqref{MEcondhausdorff1} and \eqref{MEcondhausdorff2}] read
\begin{align} \label{MEcondhausdorff1mu}
N=1:\;\;\;\; & a\leqslant \mu_1\leqslant b\,, \\ \label{MEcondhausdorff2mu}
N=2:\;\;\;\; & a^2 \leqslant \mu_1^2\leqslant \mu_2\leqslant \mu_1(b+a)-b a \leqslant b^2\,.
\end{align}
For the simplest kernel choice analogous to Eq.~\eqref{Kernel1}, $K(x)=1/(1-x)$, the condition $\mu_1^2\leqslant \mu_2$ 
yields $c_1^2\leqslant c_2 (1-\xi)$, which is not satisfied for $c_1\approx 0.3537$, $c_2\approx 0.1855$, and $\xi\approx 0.376$.
Thus, there exists no domain $D$ such that MaxEnt with the kernel $K(x)=1/(1- x)$ is applicable for the dilute Fermi gas for $N\geqslant 2$.

A kernel for which the ${N=2}$ MaxEnt moment problem can be made solvable is given by
\begin{align}
K(x)=\frac{1}{1-\alpha x}+\frac{1}{1-\beta x^2}\,.
\end{align}
For this kernel, there exists a range for $\alpha$ and $\beta$ as well as $a$ and $b$ 
for which Eq.~\eqref{MEcondhausdorff2mu} is satisfied.
However, the $N\geqslant3$ conditions are still violated.
To amend this, one may for $N=3$ add (e.g.,) a term ${\Delta K_3(x) = 1/(1-\gamma x^3)}$ to $K(x)$, 
for ${N=4}$ then (e.g.,) an additional term ${\Delta K_4(x)=1/(1+\delta x^4)}$, and so on.

There are then two possibilities to construct a sequence of MaxEnt approximants $f_{N'=1,\ldots,N}(x)$ for a given truncation order $N$, i.e.,
\begin{enumerate}[(i)]
\item one chooses $K(x)$ such that the order $N$ conditions are satisfied,
\item one uses a ``kernel scheme'' where for each subsequent $N'$ a different (i.e., extended) kernel is used.
\end{enumerate}
These two methods are however not very useful for the dilute-Fermi-gas CEP.
In both the kernel parameter space increases dimensionally with $N$ or $N'$.
A given sequence of MaxEnt approximants corresponds to a particular trajectory in this parameter space.
The dimensional increase of the parameter space with $N$ or $N'$ then makes it very difficult to identify 
approximant sequences with good convergence properties.
We therefore conclude that MaxEnt is not suitable for the dilute Fermi gas CEP.

\section{Order-dependent mapping extrapolation (ODME)} \label{sec:ODME}

\epigraph{\textit{``One extrapolation to rule them all.''}}

In this section we discuss the order-dependent mapping extrapolation (ODME) method developed in Ref.~\cite{Wellenhofer:2020ylh}.
We start by introducing a zero-dimensional (0D) model problem
in Sec.~\ref{sec:ODME1}. There, we show how the (bare) perturbation series for the 
0D model can be improved by using an order-dependent expansion point. 
Subsequently, we show that this improvement can be re-formulated as an order-dependent re-expansion of 
the bare perturbation series, corresponding to the 
``method of order-dependent mappings'' (ODM) introduced by Seznec and Zinn-Justin~\cite{Seznec:1979ev}.
In Sec.~\ref{sec:ODME2} we then state the general formulation of ODM, 
and prove that for the 0D model it converges to the exact solution.
Next, in Sec.~\ref{sec:ODME3} we introduce ODME, and show that it provides a significant improvement over ODM.
Finally, in Sec.~\ref{sec:ODME4} we discuss the application of ODME to the dilute Fermi gas CEP.

\subsection{From optimized perturbation theory to ODM} \label{sec:ODME1}

\subsubsection{Bare perturbation series}

As in Ref.~\cite{Wellenhofer:2020ylh}, we consider the partition function of the zero-dimensional $\varphi^4$ theory (see also Ref.~\cite{ZinnJustin:1980uk}):
\begin{align} \label{eq:0Dmodel}
Z(g) = \frac{1}{\sqrt{\pi}}\int_{-\infty}^\infty \!\! d\varphi \, \e^{- \varphi^2 - g \varphi^4} \,.
\end{align}
This is, of course, just an integral, whose value depends on the coupling $g$. 
We refer to it as the ``0D model''.
Its perturbative approximants are given by
\begin{align} \label{eq:PT0D}
Z_N^{\text{PT}(0)}(g) 
= \frac{1}{\sqrt{\pi}} \int_{-\infty}^\infty \!\! d\varphi \, \underbrace{\e^{- \varphi^2 } \sum_{k=0}^N\frac{(-g \varphi^4)^k}{k!} }_{\equiv z_N^{\text{PT}(0)}(\varphi,g)}
= \sum_{k=0}^N g^k \underbrace{\int_{-\infty}^\infty \!\! d\varphi \, \e^{- \varphi^2 } \frac{(-\varphi^4)^k}{\sqrt{\pi}k!}}_{\equiv c_k}\,,
\end{align}
with
\begin{align}
c_k = (-1)^k\frac{(4k)!}{2^{4k}(2k)!k!} \xrightarrow{k\rightarrow \infty} \frac{1}{\sqrt{2}\pi}(-4)^k (k-1)! \,.
\end{align}
As shown in the first block of Table~\ref{table0d}, the low-order perturbative approximants are very precise for $g\lesssim 0.01$, 
but for $g\gtrsim 0.1$ they diverge strongly from the exact result.

The divergent behavior of the perturbative approximants can be understood by comparing their integral representation with that
of the exact partition function.
The exact integrand $\e^{- \varphi^2 - g \varphi^4}$ decays smoothly with $\varphi$. The 
perturbative integrands $z_n^{\text{PT}(0)}(\varphi,g)$ on the other hand have large tails with local extrema that grow with $g$ and $N$, see Ref.~\cite{PhysRevD.57.1144}.

\subsubsection{Optimized perturbation series}

There is a straightforward way to fix this deficiency of the perturbative integrands:
one constructs an ``optimized perturbation series''
where the unperturbed partition function has as integrand $\e^{-(1+g\lambda)\varphi^2}$.
This gives
\begin{align} \label{eq:PT0Dren}
Z_N^{\text{PT}(\lambda)}(g) 
= \frac{1}{\sqrt{\pi}} \int_{-\infty}^\infty \!\! d\varphi \, 
\underbrace{\e^{- (1+g\lambda)\varphi^2 } \sum_{k=0}^N\frac{(-g \varphi^4+g\lambda\varphi^2)^k}{k!}}_{\equiv z_N^{\text{PT}(\lambda)}(\varphi,g)}\,.
\end{align}
By having the parameter $\lambda$ depend suitably on the truncation order $N$, the tails of the 
perturbative integrands can be removed, leading to an expansion whose convergence is improved over that of the bare perturbation series. We note that
this method, which we denote by ``optimized perturbation series'', is often referred to as the ``linear delta expansion'' in the 
literature~\cite{Duncan:1988hw,GANDHI1991429,PhysRevD.46.2570,PhysRevD.47.2554,PhysRevD.47.2560,PhysRevD.49.4219,PhysRevD.52.3704,GUIDA1995152,PhysRevLett.89.210403,PhysRevLett.89.271602}.
Note also that the optimized perturbation series is similar to perturbation theory
with an order-dependent reference Hamiltonian~\cite{PhysRevC.99.065811}.

One can choose $\lambda(N)$ such that the optimized perturbation series converges for all values of $g$.
As shown by Guida \textit{et al.}~\cite{GUIDA1996109}, this requires 
that ${\lambda(N)\stackrel{N\rightarrow \infty}{\sim} N^{\gamma/2}}$, with ${1\leoq \gamma<2}$.\footnote{For $\gamma=1$, the convergence condition is 
${\lambda(N)\stackrel{N\rightarrow \infty}{\sim} N^{\gamma/2}/C'}$, with $C'<C_\text{crit}$, where $C_\text{crit}\approx 1.1167$.}
In the second block of Table~\ref{table0d}, we show results for the choice ${\lambda(N)=\sqrt{N}}$.
One sees that the optimized perturbation series indeed shows convergent behavior even for large values of $g$.
Furthermore, it is more accurate than the bare perturbation series also at small $g$ where both give good results.

\subsubsection{Method of order-dependent mappings (ODM)}

The construction of the optimized perturbation series in this way may face technical difficulties in more complicated problems.
To address this, following Seznec and Zinn-Justin~\cite{Seznec:1979ev}, we now re-formulate the optimized perturbation series for the 0D model 
such that the bare perturbation series is used directly as input, i.e., no new path integrals have to be evaluated.
We first define two new parameters $\alpha$ and $w$ as follows:
\begin{align}  \label{eq:newparameters}
\alpha&= \frac{1}{\lambda+g\lambda^2},
\quad\quad
w=\frac{\sqrt{\alpha^2-4\alpha g}-\alpha}{\sqrt{\alpha^2-4\alpha g}+\alpha}\,.
\end{align}
Substituting these new parameters into the optimized series we find
\begin{align} \label{eq:ODM0D}
Z_N^{\text{PT}(\lambda)}(g)
= \sqrt{1-w} \sum_{k=0}^N w^k 
\underbrace{\int_{-\infty}^\infty \!\! d\varphi\,  \frac{1}{\sqrt{\pi}} \e^{-\varphi^2 } (\alpha \varphi^4-\varphi^2)^k}_{\equiv h_k(\alpha)} \equiv Z_N^{\text{ODM}(\alpha)}(g)\,,
\end{align}
which corresponds
to the ``method of order-dependent mappings'' (ODM).
As discussed further below in Sec.~\ref{sec:ODME2}, ODM corresponds to a truncated re-expansion (modulo a suitable prefactor)
of the bare perturbation series.

While for the 0D model, ODM and the optimized perturbation series appear to be equivalent, 
there is a difference:
in the ODM, it is $\alpha$ and not $\lambda$ that is fixed at each truncation order $N$.
(Note that the relation between $\alpha$ and $\lambda$ depends on $g$, see
Eq.~\eqref{eq:newparameters}.)
This difference between ODM and the optimized series is illustrated in Table~\ref{table0d}, 
where in the third block we show the results obtained from ODM with
$\alpha(N)=1/N$. This corresponds to ${\lambda(N)=\sqrt{N}}$ only in the limit $N\rightarrow \infty$ but not at finite $N$. 
One sees that while
the overall trend of the results is similar, for $N=3-5$,
the ODM method is slightly more precise than the optimized perturbation series both at small and large values $g$. 
However, at intermediate couplings near $g=1$ the optimized perturbation series performs better.

\subsection{ODM with fastest apparent convergence} \label{sec:ODME2}

For a function $F(x)$ with weak- and strong-coupling expansion of the form of Eq.~\eqref{eq:pertseries} and Eq.~\eqref{eq:sce},
respectively, 
the ODM approximants for the rescaled function $F(x)$ of Eq.~\eqref{fscale} are given by (with $x$ now instead of $g$)
\begin{equation}\label{eq:DOM}
f_N(x)= \xi + (1-\xi)\,[1-w(x;\alpha)]\sum_{k=0}^{N} h_k(\alpha) [w(x;\alpha)]^k \,.
\end{equation}
Here, $w(-\infty,\alpha)=1$, so the factor $(1-w)$ ensures the correct strong-coupling limit of $F(x)$: $F(\-\infty)=\xi$.
The mapping $w(x;\alpha)$ is chosen such that the (truncated) strong-coupling expansion of $F_N(x)$ matches
Eq.~\eqref{eq:sce}.
The coefficients $h_k(\alpha)$ are chosen such that 
the weak-coupling expansion of $F_N(x)$ matches Eq.~\eqref{eq:pertseries} to order $N$.
This implies that the $h_k(\alpha)$ can be
obtained by multiplying the perturbation series of the rescaled function $f(x)$
given by Eq.~\eqref{eq:scaling}
by $1/(1-w)$, substituting ${x=x(w)}$, and then expanding in powers of $w$. From this, we find
\begin{align} \label{eq:gamma_nm}
h_k=\frac{1}{(1-\xi)k!}\sum_{n,m=0}^{k} c_n \gamma_{n,m}(0)  \,,
\end{align}
with 
\begin{align}
\gamma_{n,m}(x)=  \frac{\partial^m [x(w)]^n}{\partial w^m}\bigg|_{w=w(x)}\,. 
\end{align}

The strong-coupling expansion of the 0D model 
is not of the form of Eq.~\eqref{eq:sce} but
involves fractional powers of $g$:
\begin{align}
Z(g)=g^{1/4} \sum_{k=1}^\infty  \frac{ (-1)^{k-1}}{2\sqrt{\pi}} \frac{\Gamma(k/2-1/4)}{(k-1)!}\, g^{-k/2}\,.
\end{align} 
To take this feature into account we use a prefactor $\sqrt{1-w}$ instead of $(1-w)$, 
and choose the mapping from Eq.~\eqref{eq:newparameters}.
The ODM approximants are then those given by Eq.~\eqref{eq:ODM0D} above.
The inverse mapping is given by
\begin{align}  
g(w)=\alpha\frac{w}{(1-w)^2}\,.
\end{align}
Generally, the inverse mapping will not be available in closed form. 
In that case, the coefficients $\gamma_{n,m}$ can be calculated iteratively, as detailed in the Supplementary Material of Ref.~\cite{Wellenhofer:2020ylh}.

Two commonly used prescriptions for choosing $\alpha(N)$ (or $\lambda(N)$, in optimized perturbation theory) are: 
\begin{itemize}
\item the criterion of ``fastest apparent convergence'' or ``FAC''~\cite{Seznec:1979ev,TSUTSUI2019167924} 
that fixes $\alpha(N)$ such that the order $N$ coefficient in the re-expanded series vanishes;
\item  the ``principle of minimal sensitivity`` or ``PMS''~\cite{PhysRevD.23.2916,Yukalov:2019nhu} 
meaning $\alpha(N)$ should be chosen such that the ODM approximant is least sensitive to variations of $\alpha$ about its chosen value.
\end{itemize}
Both FAC and PMS may lead to complex values of $\alpha(N)$, in which case the resulting approximant is for each $N$ defined as 
the real part of underlying ODM form.\footnote{There is also a PMS version of optimized perturbation theory (known as ``variational perturbation theory'') 
where real values of $\lambda(N)$ are enforced, see Refs.~\cite{Kleinert:2001ax,Janke:1995wt,PhysRevD.68.065001}.}

In the original 
ODM by Seznec and Zinn-Justin~\cite{Seznec:1979ev}
the parameter $\alpha(N)$ is chosen by the FAC criterion, i.e., by ${h_N=0}$. 
Adapting from Refs.~\cite{Seznec:1979ev,PhysRevD.47.2560}, we now 
show that Eq.~\eqref{eq:ODM0D} indeed converges to the correct result for ${N\rightarrow \infty}$ under the FAC choice. (A more elaborate convergence proof was provided by Guida \textit{et al.} in Ref.~\cite{GUIDA1996109}.)
From Eq.~\eqref{eq:PT0Dren}, the FAC criterion can be written as
\begin{align} \label{eq:FAClambda}
\int_{-\infty}^\infty \!\! du\,  e^{-N \phi(u)} =0\,,
\end{align}
with $u=g^2/\lambda$ and $\phi(u)=\beta u + \ln(u^2-u)$, 
where $\beta=(g\lambda^2+\lambda)/N$.
Assuming the scaling $\lambda(N)\stackrel{N\rightarrow \infty}{\sim} \sqrt{N}$ 
and $\alpha(N)\stackrel{N\rightarrow \infty}{\sim} 1/N$, respectively, 
we can evaluate Eq.~\eqref{eq:FAClambda} by steepest descent.
There are two saddle points, i.e.,
\begin{align} 
u_{1,2} =\frac{1}{2} + \frac{1}{\beta}\left(1\pm \sqrt{1+\beta^2/4}\right)\,,
\end{align}
and ${h_N=0}$ implies that $\phi(u_{1})=\phi(u_{2})$, which leads to $\beta=b_0\approx 1.325487$.
From this, we obtain
\begin{align} \label{eq:scaling}
\lambda(N)\xrightarrow{N\rightarrow \infty} \sqrt{\frac{b_0 N}{g}}\,,
\;\;\;\;
\alpha(N)\xrightarrow{N\rightarrow \infty} \frac{1}{b_0 N}\,.
\end{align}
From Eq.~\eqref{eq:PT0Dren}, the remainder $R_N(g)=Z(g)-Z_N(g)$ can be written as~\cite{PhysRevD.47.2554}
\begin{align}
R_N(g) &= \frac{\sqrt{\lambda}}{2\sqrt{\pi}} 
\int_{-\infty}^\infty \!\! \frac{du}{\sqrt{u}} 
\oint_C \frac{dz}{2\pi i}
\frac{1}{z^{N+1}(z-1)} e^{-N S(z,u)}\,,
\end{align}
where $S(z,u)=\beta u +\beta_0 z u (u-1)+\ln(z)$, 
with $u$ and $\beta$ as before, and $\beta_0=g\lambda^2/N$.
The contour $C$ encloses the poles at $z=0$ and $z=1$.
A steepest descent evaluation leads to
\begin{align}
R_N(g) &\stackrel{N\rightarrow \infty}{\simeq} \frac{\sqrt{\lambda}}{2\pi\sqrt{2 N}} 
\int_{-\infty}^\infty \!\! \frac{du}{\sqrt{u}} 
\frac{1}{z_0^{N+1}} e^{-N S(z_0,u)}\,,
\end{align}
with saddle point 
\begin{align}
z_0=\frac{1}{\beta_0u(1-u)}\xrightarrow{N\rightarrow \infty} \frac{1}{b_0u(1-u)}\,.
\end{align}
Finally, 
performing the $u$ integral by steepest descent one finds (see Ref.~\cite{PhysRevD.47.2560} for details)
\begin{align}
R_N(g) &\stackrel{N\rightarrow \infty}{\sim} N^{-3/4} e^{-\nu N}\,,
\end{align}
with $\nu\approx 0.662743$, corresponding to exponential (geometric) convergence.

The results for ODM with the FAC choice of $\alpha(N)$ are given in the fourth block of Table~\ref{table0d}.
Since the FAC criterion ${h_N=0}$ yields
several possibilities for $\alpha(N)$, we need an an additional criterion: we choose the FAC solution with smallest $h_{N-1}$. 
One sees that while the convergence pattern is qualitatively similar, FAC does yield a quantitative decrease in the errors at each order (and at each coupling) compared to what is seen with a generic $\alpha=1/N$ choice (third block of Table~\ref{tableoctic}). 
We note that we have examined ODM also with other 
$\alpha(N)$ criteria, such as different versions of PMS, but we found that they give either worse or similar results as FAC with smallest $h_{N-1}$.

\begin{table}
\caption
{Logarithmic errors $\ln[Z_N(g)-Z(g)]/\ln[10]$ of different approximants $Z_N(g)$ for the partition function of the 0D model [Eq.~\eqref{eq:0Dmodel}] 
for orders ${N=1,2,3,4,5}$ 
and ${g\in\{0.01,0.1,1.0,10\}}$. See text for details.}
\begin{center}
\begin{ruledtabular}
\begin{tabular}{l lll lllll}
\addlinespace
method 			   		   		   & $g$    		&& $N=1$	    & $N=2$ 	& $N=3$ 	& $N=4$ 	& $N=5$ 		\\
\hline
PT(${\lambda=0}$)	  			   & $0.01$ 		&&$-3.52$		&$-4.62$	    &$-5.54$ 		&$-6.35$  	&$-7.06$			\\
PT(${\lambda=0}$)	   			   & $0.1$  		&&$-1.71$		&$-1.88$		&$-1.86$		&$-1.72$		&$-1.47$			\\
PT(${\lambda=0}$)	   			   & $1$    		&&$-0.28$		&$+0.44$		&$+1.39$		&$+2.49$		&$+3.70$			\\
PT(${\lambda=0}$)	   			   & $10$   		&&$+0.85$		&$+2.51$		&$+4.43$		&$+6.51$		&$+8.72$			\\
\hline
PT(${\lambda=\!\!\sqrt{N}}$)	   & $0.01$ 		&&$-3.93$		&$-5.12$ 		&$-6.12$		&$-7.01$ 		&$-7.79$			\\
PT(${\lambda=\!\!\sqrt{N}}$)	   & $0.1$  		&&$-2.3$		&$-2.81$ 		&$-3.07$		&$-3.22$		&$-3.29$			\\
PT(${\lambda=\!\!\sqrt{N}}$)	   & $1$    		&&$-1.64$		&$-2.25$ 		&$-2.64$		&$-3.20$		&$-3.53$			\\
PT(${\lambda=\!\!\sqrt{N}}$)	   & $10$   		&&$-0.82$		&$-0.94$		&$-1.04$		&$-1.14$		&$-1.22$			\\
\hline
ODM(${\alpha=1/N}$)		 		   & $0.01$ 		&&$-3.80$		&$-5.19$		&$-6.42$		&$-7.53$		&$-8.57$			\\
ODM(${\alpha=1/N}$)		 		   & $0.1$  		&&$-2.13$		&$-2.80$		&$-3.34$		&$-3.83$		&$-4.27$			\\
ODM(${\alpha=1/N}$)		 		   & $1$    		&&$-1.21$		&$-1.61$		&$-1.89$		&$-2.16$		&$-2.39$			\\
ODM(${\alpha=1/N}$)		 		   & $10$   		&&$-1.02$		&$-1.34$		&$-1.55$		&$-1.75$		&$-1.92$			\\
\hline
ODM(FAC)						   & $0.01$ 		&&$-3.94$		&$-5.31$		&$-6.40$		&$-7.32$		&$-8.89$			\\
ODM(FAC)						   & $0.1$  		&&$-2.32$		&$-2.99$		&$-3.45$		&$-3.81$		&$-4.85$			\\
ODM(FAC)						   & $1$    		&&$-1.50$		&$-1.95$		&$-2.25$		&$-2.49$		&$-3.23$			\\
ODM(FAC)						   & $10$   		&&$-1.36$		&$-1.75$		&$-2.01$		&$-2.22$		&$-2.85$			\\
\hline
ODME							   & $0.01$ 		&&$-4.14$		&$-5.36$		&$-6.98$		&$-8.04$		&$-9.37$			\\	
ODME							   & $0.1$  		&&$-2.64$		&$-3.19$		&$-4.23$		&$-4.73$		&$-5.53$			\\
ODME							   & $1$    		&&$-2.20$		&$-2.58$		&$-3.32$		&$-3.68$		&$-4.26$			\\
ODME							   & $10$   		&&$-2.54$		&$-2.88$		&$-3.51$		&$-3.83$		&$-4.33$			\\
\end{tabular}
\end{ruledtabular}
\end{center}
\label{table0d}
\end{table}

\subsection{From ODM to ODME} \label{sec:ODME3}

Evidently, the choice of the criterion to fix the mapping parameter $\alpha(N)$ plays 
an important part in the application of ODM.
We have seen that for the 0D model the FAC criterion improves upon the simple scaling choice $\alpha(N)=1/N$.
However, the FAC (or PMS) choice appears somewhat heuristic, and it seems plausible that choices of $\alpha(N)$ which give better results should exist.

A second issue is that in the CEP we want to include strong-coupling information beyond the unitary limit.
One possible method for this was proposed (for optimized perturbation theory) by Kleinert in Ref.~\cite{KLEINERT1995133}.
However, that method still leaves the first issue, the choice of $\alpha(N)$.

The ``order-dependent-mapping extrapolation" (ODME) approach introduced in Ref.~\cite{Wellenhofer:2020ylh} tackles these two issues at once. 
That is, in the ODME 
one fixes $\alpha(N)$ by ensuring that the strong-coupling expansion of $F_N(x)$ has a first-order coefficient equal to the first correction to the strong-coupling limit, i.e., $d_1$ of Eq.~(\ref{eq:sce}).
This again yields several possibilities for $\alpha(N)$; we select the one that minimizes 
the difference between $d_2$ and the corresponding coefficient of $F_N(x)$.

The ODME results for the 0D model are shown in the fifth block of Table~\ref{table0d}.
One sees that the ODME choice for $\alpha(N)$ leads to much more precise approximants than FAC.
In particular, ODME is more precise also at very small values of $g$ where including additional strong-coupling information is not expected to give improved results.
Further, in Ref.~\cite{Wellenhofer:2020ylh} we showed that at low truncation orders ODME 
outperforms also the strong-coupling expansion (which has infinite radius of convergence), even at large $g$.
For example, for ${g=10}$ and $N\leqslant 4$ ODME is more accurate than the strong-coupling expansion truncated at the same order, and
even at ${g=100}$ they reach similar precision for $N=4$.
Further, at ${g=1}$ the strong-coupling expansion reaches a higher precision than ODME not until $N\geqslant 14$. Not surprisingly, 
in the weak-coupling regime ($g\lesssim 1$) the strong-coupling expansion produces accurate results only for very large truncation order.

The ODME approach of Ref.~\cite{Wellenhofer:2020ylh} makes also constructive use of the mapping ambiguity of ODM: 
among the available mappings, in ODME one selects those that give the best converged results at low orders.
As a measure for convergence 
we consider the sum of the 
deviations of consecutive-order approximants:
\begin{align} \label{eq:errorODME}
\Delta_w = \sum_{x_i} {\sum_{N=2}^M \sigma_N |F_{N}(x_i)-F_{N-1}(x_i)|}\,.
\end{align}
Here, $\{x_i\}$ are selected points in the coupling regime of interest, and $M$ is 
the truncation order up to which the perturbation series is known.
The weights $\sigma_N$ may be chosen in accord with the principle that the deviation should be smaller at higher orders, e.g., one may set ${\sigma_N=N}$.
In the ODME, 
one chooses the mappings $w(x)$ for which the quantity $\Delta_w$ is smallest.
We will make use of this criterion of low-order convergence when we apply ODME to the dilute Fermi gas CEP.

\subsection{Application to dilute Fermi gas} \label{sec:ODME4}

We now discuss the application of ODME for the 3D Fermi gas. 
We consider mappings of the form
\begin{align}
    w(x)=-w_0 \frac{x}{D(x;\alpha)}\,,
\end{align}
with ${w_0=\lim\limits_{x\to-\infty}D(x;\alpha)/x}$. A general form for $D(x;\alpha)$
consistent with the SCE of the 1D and 3D Fermi gas is, e.g.,
\begin{align}
	D(x;\alpha)=\kappa_{1}\alpha-\kappa_{2}x +(\kappa_{3}\alpha^{\mu}+(-x)^{\nu})^{1/{\nu}}\,.
\end{align} 
In principle large sums of such terms could be used. However, we found that in order to have
well-converged results at low orders, excessively complicated forms of $D(x;\alpha)$ are disfavored.

\begin{figure*}[b] 
\centering
\includegraphics[width=0.81\textwidth]{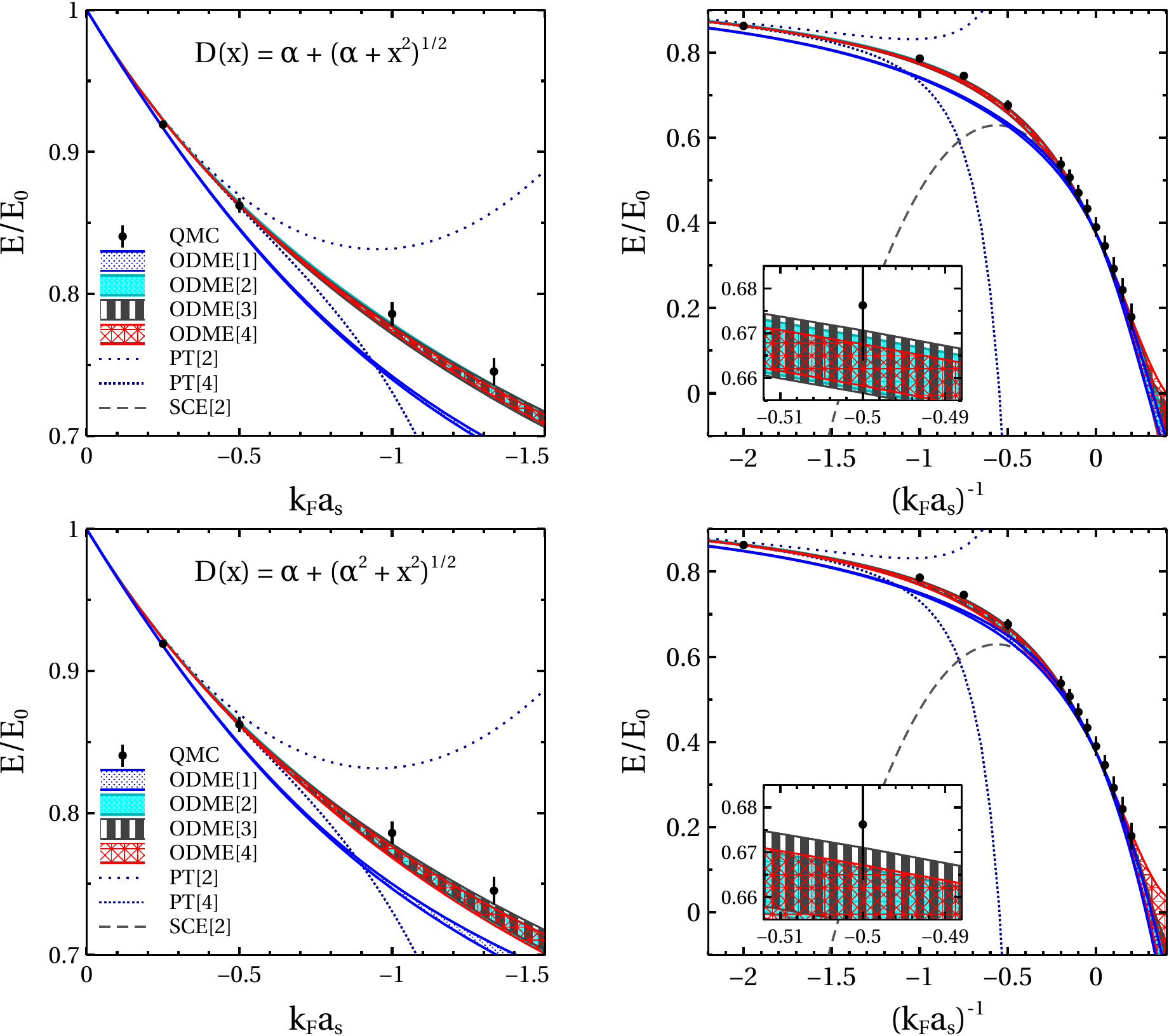} 
\caption{ODME results for the dilute Fermi gas for two different mappings. See text for details.}
\label{fig:ODMEplot}
\end{figure*}

In Ref.~\cite{Wellenhofer:2020ylh} we examined several choices for $D(x;\alpha)$.
We found a number of mappings that give quite similar results in the BCS regime.
Among them, the two mappings for which the ODME approximants are best converged have
\begin{align} \label{eq:map1}
	D(x;\alpha)=\alpha + \sqrt{\alpha + x^2}\,,
\end{align} 
and 
\begin{align} \label{eq:map2}
	D(x;\alpha)=\alpha + \sqrt{\alpha^2 + x^2}\,.
\end{align} 
Of the mappings we surveyed the quantity $\Delta_w$ of Eq.~\eqref{eq:errorODME} 
is smallest for these two (we chose $x_i$ evenly distributed in the BCS regime).

The results for the dilute Fermi gas CEP from these two mappings are shown in Fig.~\ref{fig:ODMEplot}.
The bands correspond to the uncertainties in the values 
${d_1=-0.90(5)}$ and ${d_2=-0.8(1)}$. We used the central value of the experimental result for the Bertsch parameter, ${\xi=0.376}$.
For both mappings the ODME results are below the central (variational)
QMC points, and the ODME bands overlap with the QMC error bars for $N\geqslant 2$. 
While the first-order ($N=1$) ODME results are closer to the QMC points for the second mapping [Eq.\eqref{eq:map2}], 
the ODME bands are smaller for the first mapping, corresponding to a slightly smaller $\Delta_w$ (averaged over the $d_1$ and $d_2$ input).

The ODME approximants can be easily continued into the BEC region, and as seen in Fig.~\ref{fig:ODMEplot},  
they still give reasonaly well converged results for BEC values of $x$ that are not too far from the unitary limit.
For both mappings the dependence on the strong-coupling input in the BEC region is larger for $N=4$.
This may however be improved by also including BEC values into $\Delta_w$
and by selecting sequences of ODME approximants according to their 
convergence for each ${(\xi,d_1,d_2)}$ input value.

With approximant sequences that have good convergence properties in hand, we can 
use sequence extrapolation techniques such as the Shanks transformation to obtain estimates 
for $N\rightarrow\infty$ results.
For example, for ${x=-2}$ and the central values of $d_1$ and $d_2$ quoted above, 
the second mapping [Eq.\eqref{eq:map2}] gives the ODME values 
\begin{align}
F_{N}(-2)\approx (0.644, 0.660, 0.663, 0.665)
\end{align}
for ${N=(1,2,3,4)}$, which 
approaches the QMC value ${F_\text{QMC}(-2)= 0.676(12)}$.
The Shanks extrapolated result is ${F_{\infty}(-2)} \approx 0.670$, which is well within the QMC error bar.
For the first mapping [Eq.\eqref{eq:map2}] we get instead 
\begin{align}
F_{N}(-2)\approx (0.631, 0.664, 0.663, 0.663)
\end{align}
for ${N=(1,2,3,4)}$.
The Shanks transformation gives ${F_{\infty}(-2)} \approx 0.663$, which is just below the QMC error bar.
Larger values of $d_1$ generally give results that are closer to the QMC data, e.g., for $d_1=0.95$ 
one gets 
\begin{align}
F_{N}(-2)\approx (0.634, 0.669, 0.670, 0.668)
\end{align}
and ${F_{\infty}(-2)} \approx 0.670$.
Note also that the QMC point at unitarity is ${F_\text{QMC}(-\infty)= 0.390(18)}$, i.e., 
the experimental Bertsch parameter ${\xi=0.376(4)}$ lies at the lower end of the QMC error bar at unitarity.
This perhaps provides a hint that the exact result at ${x=-2}$ is actually closer to the ODME results than to the central QMC value used here.

\section{Summary and Outlook} \label{sec:conclusion}

\epigraph{\textit{``Don't adventures ever have an end? I suppose not. Someone else always has to carry on the story."}\\ Bilbo Baggins}

The problem of extrapolating a perturbation series beyond the weak-coupling regime is ubiquitous in theoretical physics.
The CEP
is to find sequences of perturbation-series extrapolants 
that take into account limited strong-coupling information and are well converged at low orders. The strong-coupling information can come either from experiment of from numerical simulations. 
In this paper we investigated several different approaches to the CEP. 
As our target problem we focused on the constrained extrapolation of the perturbation series for the ground-state energy of the dilute Fermi gas throughout the entire BCS regime and into the BEC regime.

First, we examined two standard extrapolation methods, Pad\'e and Borel resummation.
We found that the results from these methods for our target problem are deficient in significant ways.
While Pad\'e is a straightforward approach and can implement any number of SCE constraints, it is unreliable because unphysical poles can appear in the extrapolation region. 
The more sophisticated Borel method
does not produce spurious poles 
and reliably improves the convergence of perturbative results in the weak-coupling region.
However, we found that for many Borel extrapolants
the convergence breaks down at intermediate-to-strong coupling (see also Ref.~\cite{Wellenhofer:2021eis}),
even if they are constrained by strong-coupling information. 
Moreover, because they are based on an integral transform, Borel extrapolants cannot be easily continued into the BEC regime.

We then turned to the more flexible MaxEnt extrapolation technique, 
which can provide better results than standard resummation methods in certain cases. However,
its application to the dilute Fermi gas is inhibited by the fact that the moment conditions, which must be satisfied for it to be applicable, 
require quite intricate choices
for the MaxEnt kernel. This renders the kernel highly ambiguous.

Finally, we introduced the ODME method developed in Ref.~\cite{Wellenhofer:2020ylh}, building on previous work by Seznec and Zinn-Justin~\cite{Seznec:1979ev}.
Instead of matching the low-order weak-coupling coefficients to an integral form (Borel, MaxEnt) or a fixed functional form (Pad\'e) this method uses a re-expansion of the perturbation series in terms of a mapping $w_\alpha(x)$.
Strong-coupling constraints up to second order are implemented by adjusting the (truncation-order-dependent) mapping parameter $\alpha(N)$.
Extrapolation into the BEC region is as straightforward as in the Pad\'e case.
We found that ODME outperforms similar methods for a 0D model, and 
we showed that it provides good results for the dilute Fermi gas CEP. 
In particular, simple mappings already produce good results and reasonable variations of the mapping choice do not significantly affect the obtained results.
Overall, compared to the standard Pad\'e and Borel methods as well as the MaxEnt technique, 
ODME is more generally applicable
and does not induce artifacts such as spurious poles. 
Moreover, our results suggest
that, already at low truncation orders, ODME often leads to well-converged results all the way from weak to strong coupling.

The ODME method is very flexible, and can be applied to a variety of extrapolation problems.
In particular, this method can also be applied in cases 
where the available weak- and strong-coupling data does not provide sufficient constraints, but 
additional information at an intermediate coupling is available.
That is, by mapping the
intermediate coupling point to infinity via a conformal transformation, 
ODME can be used to construct weak-to-intermediate coupling approximants, which can 
then in a second application of the ODME method be analytically continued to the strong-coupling regime.
An interesting target for future applications of ODME along these lines is the construction of an equation of state of strongly interacting matter in neutron stars that combines constraints from low-density nuclear physics calculations, observational constraints at higher density, and ultra-high-density perturbative QCD (see, e.g., Ref.~\cite{Huth:2020ozf}).

\acknowledgements
We thank R.~J.~Furnstahl, S. Gandolfi, and A. Gezerlis for useful discussions.
DRP is grateful for the warm hospitality of the IKP Theoriezentrum Darmstadt. 
He is also grateful for twenty years of friendship and stimulating discussions about science, Roman coinage, the Lord of the Rings, and disparate other topics with Dist.~Prof.~D.~A. Drabold. 
This research was supported by the Deutsche Forschungsgemeinschaft (DFG, German Research Foundation) -- Project-ID 279384907 -- SFB 1245, the US Department of Energy (contract DE-FG02-93ER40756), and by the ExtreMe Matter Institute.

\end{document}